\documentclass{aa}  

\usepackage{graphicx}
\usepackage{txfonts}
\usepackage{lipsum}
\usepackage{subcaption}         
\usepackage{lscape}             
\usepackage{placeins}           
                                
\usepackage{graphicx}
\usepackage{appendix}
\usepackage{subcaption}

\usepackage{txfonts}

\usepackage{color}
\usepackage{natbib,twoopt}
\usepackage[breaklinks]{hyperref}      
\bibpunct{(}{)}{;}{a}{}{,}             
\hypersetup{
  colorlinks,
  citecolor=blue,
  linkcolor=blue,
  urlcolor=blue,
}
\makeatletter
  \newcommandtwoopt{\citeads}[3][][]{\href{http://adsabs.harvard.edu/abs/#3}%
    {\def\hyper@linkstart##1##2{}%
     \let\hyper@linkend\@empty\citealp[#1][#2]{#3}}}
  \newcommandtwoopt{\citepads}[3][][]{\href{http://adsabs.harvard.edu/abs/#3}%
    {\def\hyper@linkstart##1##2{}%
     \let\hyper@linkend\@empty\citep[#1][#2]{#3}}}
  \newcommandtwoopt{\citetads}[3][][]{\href{http://adsabs.harvard.edu/abs/#3}%
    {\def\hyper@linkstart##1##2{}%
     \let\hyper@linkend\@empty\citet[#1][#2]{#3}}}
  \newcommandtwoopt{\citeyearads}[3][][]%
    {\href{http://adsabs.harvard.edu/abs/#3}
    {\def\hyper@linkstart##1##2{}%
     \let\hyper@linkend\@empty\citeyear[#1][#2]{#3}}}
\makeatother
\definecolor{cobalt}{rgb}{0.06, 0.2, 0.65}

\definecolor{smalt(darkpowderblue)}{rgb}{0.0, 0.2, 0.6}
\definecolor{forestgreen(traditional)}{rgb}{0.0, 0.5, 0.0}

\newcommand{\Msun}{M$_{\odot}$}
\newcommand{\Mjup}{M$_{\mathrm{J}}$}

\defcitealias{mauchsorianoetal26-1}{Paper\,I}

\defcitealias{Villaver_2009}{VL09}
\defcitealias{rasioetal96-1}{R96}
\usepackage[normalem]{ulem}

\begin{document}

   \title{Predicted JWST yields of giant planets around white dwarfs} 


%
%
%
   \titlerunning{Predicted JWST yields of giant planets around white dwarfs}

   \author{Alex Mauch-Soriano\inst{1,2}\fnmsep\thanks{Corresponding author: alex.mauch@usm.cl},
          Matthias R. Schreiber
          \inst{1}
          \and
          Maria Paula Ronco
          \inst{3}
          \and
           Felipe Lagos-Vilches\inst{1}
          \and
          Diogo Belloni 
          \inst{4}
          \and
          Wolfgang Brandner\inst{5}
          }

   \institute{Departamento de F\'isica, Universidad T\'ecnica Federico Santa Mar\'ia, Av. Espa\~na 1680, Valpara\'iso, Chile
         \and
         {Instituto de F\'isica, Pontificia Universidad Cat\'olica de Valpara\'iso, Casilla 4059, Valpara\'iso, Chile}
            \and
            {Instituto de Astrof\'{\i}sica de La Plata, CCT La Plata-CONICET-UNLP, Paseo del Bosque S/N (1900), La Plata, Argentina}
                \and
                 International Centre of Supernovae (ICESUN), Yunnan Key Laboratory of Supernova Research, Yunnan Observatories, CAS, Kunming 650216, China
                \and
                Max Planck Institute for Astronomy, Königstuhl 17, 69117 Heidelberg, Germany
                }

   \date{Received September 30, 20XX}

 
  \abstract
   {Surveying substellar companions to white dwarfs (WDs) may provide constraints on planet populations that are hard to access around main-sequence stars, thereby informing planet formation theories. For WD progenitors in the $1-4$~\Msun~ mass range, the occurrence rates and detectability of these companions can be predicted using results from radial velocity surveys of red giant stars.}
   {Building on our previous population synthesis, we here aim to determine the detectability with the James Webb Space Telescope (JWST) of substellar objects around WDs that descended from stars with masses of $1-4$~\Msun.} 
   {We derive the radii and temperatures of the predicted substellar companions and their host stars.
   Combined with the technical specifications of JWST, this provides luminosities, fluxes, and contrast ratios which can then be used to construct detectability maps. Based on these maps, we estimate the probability of detecting companions either through direct imaging or infrared excess.  }
   {We find that around 20 to 35 \% of the predicted companions within $30$~pc can be detected either through direct imaging or infrared excess at wavelengths $\gtrsim15\,\mu$m and at a significance level of 3--5$\sigma$. 
   Applying these results to the observed WD population within this distance, we predict up to one detectable substellar companion.
   Assuming we dramatically overestimated tidal forces, this number increases to a few. Accounting for the relatively unconstrained potential population of substellar companions at large separations indicated by imaging surveys, the number of detectable substellar companions could reach nearly 20.}
   {We predict JWST to detect between zero and nearly 20 planetary companions if all WDs that descended from stars with masses of $1-4$~\Msun~within $30$~pc are observed. Searching for wide substellar companions around more massive WDs might be more promising. 
   If companions are formed through gravitational instabilities around their massive progenitors, these companions migrate further outward during post-main-sequence mass loss, placing them at orbital separations easily probed by JWST. }


   \maketitle

\section{Introduction}

The population of substellar companions to white dwarfs (WDs) offers the potential to constrain planet formation around 
a large range of stellar masses. The current population of WDs in the Milky Way formed from $\sim1-8$~\Msun~main sequence stars. 
Giant planet formation through core accretion is expected to dominate for stars less massive than $\sim4$~\Msun, although fragmentation of gravitationally unstable disks may produce an additional population of substellar companions at large orbital separations.
For more massive stars, the lifetimes of protoplanetary disks are probably too short for efficient giant planet formation through core accretion \citep{Kunitomo2021,Komaki2021,roncoetal24-1}. Substellar companions at large 
separations around $4-8$~\Msun~stars are therefore likely to have formed through fragmentation of gravitationally unstable discs \citep{Kratter2016,Longarini2026}. 
These stellar mass ranges translate to WD masses from $\sim0.5-0.85$~\Msun~for progenitor stars with masses of $1-4$~\Msun~and $0.85-1.4$ for more massive stars up to $\sim8$~\Msun~\citep{Cummingsetal2018,marigo2020}.

Searching for substellar companions around WDs may therefore provide unique information on different planet formation theories.  
Considerable observational efforts have been devoted to detecting this substellar companion population. Several surveys have used direct-imaging techniques to search for massive giant planets around WDs, but these searches have so far produced few convincing detections \citep[][Lagos-Vilches et al. A\&A~submitted]{burleighetal02-1,debesetal05-1,hoganetal09-1,brandneretal21-1}. These imaging campaigns have been complemented by transit searches \citep{van-sluijsetal18-1} and searches for infrared (IR) excesses \citep{farihietal08-1}. Nevertheless, only a small number of substellar companions to WDs are currently known.

Some of these objects, including the proposed Neptune-like planet around WD\,J0914+1914 \citep{gaensickeetal19-1} and the transiting Jupiter-sized planet around WD\,1856+534 \citep{vandenburgetal20-1}, orbit their WDs at very short periods. Their present-day configurations therefore require substantial orbital evolution, potentially driven by interactions with additional substellar companions or by survival of a common-envelope phase \citep{munoz+petrovic20-1,lagosetal21-1,maldonadoetal21-1,Chamandy-21,Stephan-21,mirandaetal2026}. 

Other candidates are found at much wider separations \citep[e.g.,][]{luhmanetal12-1, blackmanetal21-1}, where orbital expansion in response to stellar mass loss may have played an important role in shaping their present-day orbits. However, the evolutionary origins of several individual systems remain uncertain, and the currently known objects do not yet provide a robust empirical census of the planetary population around WDs.

The small number of detections may largely reflect the limitations of previous surveys rather than an intrinsic absence of planets \citep{brandneretal21-1}. WDs are intrinsically faint, which provides favourable planet-to-star contrasts, but the expected substellar companions are generally old, cool, and correspondingly difficult to detect. The detectability of a surviving planet therefore depends not only on its mass and orbital separation, but also on the age and mass of the WD, the age and distance of the system, and the wavelength coverage, sensitivity, and angular resolution of the observing facility. The advent of the James Webb Space Telescope (JWST) has substantially improved the prospects for detecting these objects. 

The unprecedented mid-IR sensitivity of JWST has recently opened a new window for detecting giant planets around WDs.
The capabilities of JWST are most convincingly illustrated by WD 1856+534\,b. Mid-Infrared Instrument \citep[MIRI, e.g.][]{Wright2023} observations detected its thermal emission and confirmed its planetary mass, while subsequent Near Infrared Spectrograph (NIRSpec) transmission spectroscopy revealed a methane-rich, aerosol-bearing atmosphere and constrained its mass and temperature \citep{limbachetal25-1,macdonaldetal26-1} and thereby the origin of the close orbit. JWST can therefore not only detect mature giant planets around WDs but, in favourable systems, also characterize their atmospheres and constrain their post-main-sequence dynamical histories. 

However, despite ongoing surveys, a substantial population of planetary companions around WDs has yet to be established.
MIRI observations have revealed several spatially unresolved IR excesses consistent with cold Jovian planets, including candidates around WD\,0310$-$688, WD\,2105$-$820, WD\,2149+021, and WD\,0644+025 \citep{limbachetal24-1,debesetal25-1, poulsenetal26-1}. If planetary, these companions have inferred masses of a few Jupiter masses and orbital separations ranging from less than 1\,au to approximately 10\,au. However, unusual cold debris disks can reproduce at least some of the excesses. Two initially resolved planet candidates were subsequently identified as background galaxies through second-epoch imaging \citep{mullallyetal24-1,mullallyetal26-1}. More recently, a spatially resolved cold brown dwarf candidate, named MEAD 62\,b, was found around 2MASS\,J09424023$-$4637176, a nearby WD. However, the red color of this candidate substellar companion is similar to that of background galaxies, making it equally likely to be a false positive or a real substellar companion to the  WD \citep{albertetal26}.   

These results demonstrate both JWST’s exceptional sensitivity to old, cold planets and the importance of follow-up observations for establishing their planetary nature. 
Some current programs are dedicated to confirming or disproving candidates from previous cycles, for example GO $12573$ \citep{Vanderburg2026_JWST} and GO $11868$ \citep{Poulsen2026_JWST},  while program GO $9579$ \citep{Mullally2026_JWST} stands out as a high-sensitivity search through direct imaging, reaching planet masses below half a Jupiter mass for every target. 
Despite all these efforts, so far not a single planet around a WD has been discovered and confirmed by ongoing JWST programs. 

To understand this lack of detections, theoretical predictions are essential.  
In \citet[][hereafter Paper\,I]{mauchsorianoetal26-1}, we developed a population-synthesis model to predict the occurrence rate and present-day properties of substellar companions around WDs descending from stars less massive than $4$\,\Msun. 

We found a low occurrence rate of $3\pm1.5$ \% that depends sensitively on the adopted age--metallicity relation, but does not exceed $\sim8$ \%, even in the most favourable case. In \citet{mirandaetal2026} we also estimated the fraction of substellar companions surviving common-envelope evolution and found that the occurrence rate of such survivors is negligible. 

These low intrinsic occurrence rates already provide a first indication of why most surveys have not detected large numbers of substellar companions to WDs with masses between $0.5$ and $0.85$~\Msun. 
However, a direct comparison between these predictions and the results of observational surveys requires modelling the present-day luminosities and projected separations of the substellar companions and applying realistic instrumental detection limits.

In this second paper of the series, we therefore investigate the detectability of the surviving substellar companions predicted by our population synthesis. We determine how the probability of detection depends on the properties of the WD, the substellar companion, and the observing facility, and quantify the fraction of the underlying population accessible to current and future direct-imaging surveys. We thereby translate the intrinsic population predicted in \citetalias{mauchsorianoetal26-1} into observationally testable detection rates, identify the most promising WD targets, and assess the potential of JWST to establish the first statistically meaningful census of giant planets around WDs with masses below $\sim0.85$~\Msun.

\section{Stellar and substellar companion properties and detection limits}

In \citetalias[][]{mauchsorianoetal26-1}, we predicted the population of substellar companions around WDs descending from stars with initial masses of $1$--$4$\,\Msun. We generated a population of WD progenitors following the initial mass function of \citet{kroupa2001}, assigning each star an age assuming a constant star formation rate \citep[e.g.][]{Cukanovaite2023} and a metallicity according to the age--metallicity relation of \citet{carrillo2023}. We then assigned substellar companions based on the occurrence rates inferred by \citet{wolthoff2022}, including their dependence on stellar mass and metallicity. Finally, we followed the stellar and orbital evolution of these systems using detailed evolutionary tracks computed with MESA \citep[e.g.][]{Paxton2019} in combination with FATES \citep{roncoetal20-1}, thereby predicting the present-day population of substellar companions around WDs descending from main-sequence stars less massive than $4$\Msun.

The main result of that study was a total substellar companion occurrence rate of $3\pm1.5$ \% around WDs descended from $1$--$4$\,\Msun~stars. We also derived a probability map describing the dependence of the substellar companion occurrence rate on WD mass and total system age. The predicted population is concentrated primarily around relatively young systems, with total ages of approximately $1$--$6$\,Gyr, and comparatively low-mass WDs \citepalias[see][figure 5]{mauchsorianoetal26-1}.

In what follows, we combine these predicted occurrence rates with substellar companion detectability models to determine the observable fraction of this population.
More specifically, we investigated detection probabilities taking into account 
the capabilities of JWST to detect either an IR excess or a spatially resolved substellar companion.

Evaluating the detectability of our predicted substellar companion population therefore requires not only the substellar companion masses and orbital separations, but also the present-day spectral energy distributions of both components.

\subsection{White dwarf and substellar companion radii and temperatures}

To determine the radii, effective temperatures, and surface gravities of the WDs, we interpolated the cooling tracks of \citet{bedardetal2020} using the WD masses and cooling ages predicted by our population synthesis. We assumed that all WDs have hydrogen-rich atmospheres. We then modelled their spectral energy distributions using the open-source package \texttt{species} \citep{Stolkeretal2020}, adopting synthetic WD spectra based on the atmosphere models of \citet{Koester2010}. Synthetic photometry was subsequently calculated for the relevant JWST filters, yielding the absolute magnitude of each WD in each band. The \texttt{species} package includes the corresponding filter-transmission curves, retrieved from the SVO Filter Profile Service \citep{Rodrigo2020}, allowing the model spectra to be integrated directly over each instrumental passband.

For the substellar companions, we adopted the Sonora Flame Skimmer models\footnote{\url{https://zenodo.org/records/20030439}} \citep{Mang_2026}, the latest generation of cloud-free, one-dimensional atmospheric and evolutionary models.  Specifically, we used the chemical-equilibrium models assuming a carbon-to-oxygen ratio of $\mathrm{C/O}=0.5$ and a metallicity of $[\mathrm{M/H}]=0.5$. The adopted C/O ratio is close to the solar value, while the enhanced metallicity was chosen to approximately account for the metal enrichment observed in Jupiter relative to the Sun \citep[see, e.g.,][and references therein]{Lodders2010, Guillot2015}. These models extend the available parameter space to effective temperatures as low as $\sim50$\,K. For each combination of substellar companion mass and total system age, we interpolated the evolutionary tables to obtain the present-day radius, effective temperature, and surface gravity. We then determined the absolute magnitudes of the substellar companion in the relevant JWST filters by interpolating the synthetic-photometry tables provided by the models. Six of the 262 substellar companions from our population synthesis had ages exceeding the maximum age covered by the evolutionary tracks corresponding to their masses. Since their magnitudes could not be reliably estimated, we conservatively treated these companions as undetectable in our analysis.

\subsection{Detection limits}

Using the synthetic photometry and physical properties of both components, we evaluated the detectability of each predicted WD–substellar companion system. For each substellar companion, the projected angular separation determines whether it can be spatially resolved or must instead be detected through its IR excess. In either case, detectability depends on the substellar companion's flux in the corresponding filter. We calculated projected separations from the predicted semimajor axes and eccentricities by assigning random orbital inclinations and phases based on the formalism implemented in \texttt{Exo-DMC} \citep{Bonavita2020}. For each companion, we generated 1000 orbital realizations with isotropic orientations (uniformly distributed $\cos i$, longitude of the ascending node $\Omega$, and argument of periastron $\omega$) and mean anomalies uniformly distributed over $[0,2\pi)$, i.e., sampling the orbital phase uniformly in time. After solving Kepler's equation, the orbital positions were projected onto the sky plane using the Thiele--Innes constants and converted into angular separations for a given distance. These separations and the companion magnitudes were then compared with the filter- and exposure-time-dependent JWST detection limits.

For our analysis, we consider four different MIRI filters (F1000W, F1500W, F1800W, and F2100W), which have been used by the largest JWST surveys for giant planets around WDs to date, i.e., the MIRI survey for Exoplanets Orbiting White-dwarfs \citep[MEOW,][]{limbachetal24-1}, which observed 17 WDs using F1800W and F2100W and the MIRI Excess Around Degenerates Survey \citep[MEAD,][]{albertetal26}, which targeted 56 different WDs through F1000W and F1500W. Both surveys searched for Jupiter-like companions, reaching the super-Jupiter mass regime through IR excess and direct imaging. To define our detection thresholds, we considered exposure times informed by the observational setups of these surveys for the corresponding filters, with exposure times of $t=444$~s for both F1000W and F1500W, and $t=277.5$~s and $t=710$~s for F1800W and F2100W, respectively. These setups correspond to 5$\sigma$ detection limits of approximately $2.7\,\mu$Jy for F1000W and F1500W, and $6.6\,\mu$Jy for F1800W and F2100W, according to the JWST Exposure Time Calculator\footnote{\url{https://jwst.etc.stsci.edu/}} assuming medium background noise.

While JWST’s spatial resolution is primarily determined by its wavelength-dependent angular resolution, its ability to resolve a planet from its host WD is further limited by the brightness contrast between the WD and the planet, where bright stellar diffraction wings and residual point spread function (PSF) subtraction noise overwhelm the planet's faint thermal emission at small angular separations.
We consider a substellar companion to be spatially resolved if the angular separation exceeds $1.5\times \text{FWHM}$ (full width at half maximum) for the corresponding filter, adopting the filter-dependent FWHM values reported by \citet{Dicken_2024}. This provides a conservative criterion for reaching the necessary contrast for Jovian planets that are a few Gyr old \citep{Poulsenetal2024}. 

Substellar companions that remain spatially unresolved may instead be detected through the IR excess they produce. In this approach, a bluer filter, in which the contribution from the substellar companion is minor, is used to anchor the WD atmosphere model. The predicted WD flux is then compared with the combined WD-plus-substellar companion flux in a redder filter, where even a cool substellar companion could contribute significantly. 
Given that the absolute photometric calibration uncertainty of MIRI is approximately $2\%$

of the measured flux value, an excess of approximately $6$--$10$\% in the total system flux is generally required for $3\sigma$ and $5\sigma$ detections, respectively \citep{debesetal25-1,Gordon_2022,Gordon_2025}. We estimate the statistical significance considering $\sigma = \sqrt{((F_\text{WD}+F_\text{SC})\times0.02)^{2}+\sigma_{\text{phot}}^{2}}$, where $F_\text{WD}$ and $F_\text{SC}$ denote the fluxes of the white dwarf and substellar companion, respectively, and $\sigma_{\text{phot}}$ represents the expected $1\sigma$ photometric sensitivity of the corresponding MIRI filter for the adopted exposure time, which we use as an estimate of the statistical photometric uncertainty.

\section{Detection probabilities}

\begin{figure*}
    \centering
    \begin{minipage}{1.0\textwidth}
        \centering
        \includegraphics[width=0.99\textwidth]{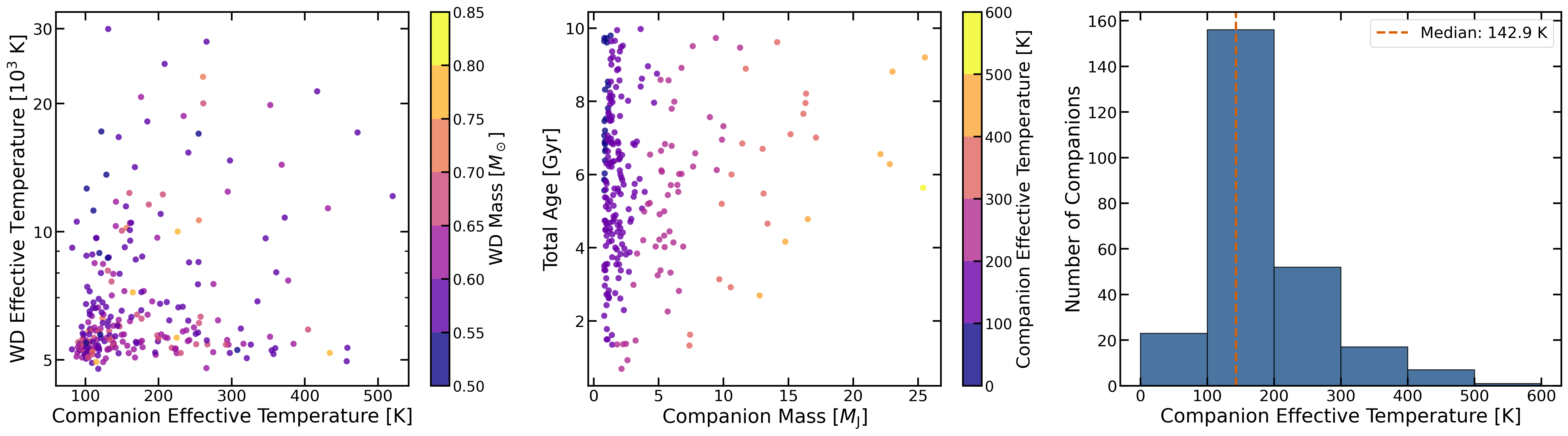}  
    \end{minipage}
    \caption{The effective temperatures of the predicted WDs and their substellar companions from our population synthesis (left). Most WDs are rather cold ($\lesssim6000$~K) and the substellar companion temperatures cluster around low temperatures as well ($\lesssim150$~K). These cool temperatures are caused by the total age of the systems and the masses of the companions ($\sim1$~\Mjup, middle panel). The median value of the temperature distribution of the companions is $142$~K (right panel). }
    \label{fig:twd_ssc}
\end{figure*}

In what follows, we first present the detection probability for the substellar companions predicted by our population synthesis before we apply our results to WDs in the solar neighborhood. In both cases we consider detections through direct imaging and IR excess.

\begin{figure*}
    \centering
    \begin{minipage}{1.0\textwidth}
        \centering
        \includegraphics[width=0.99\textwidth]{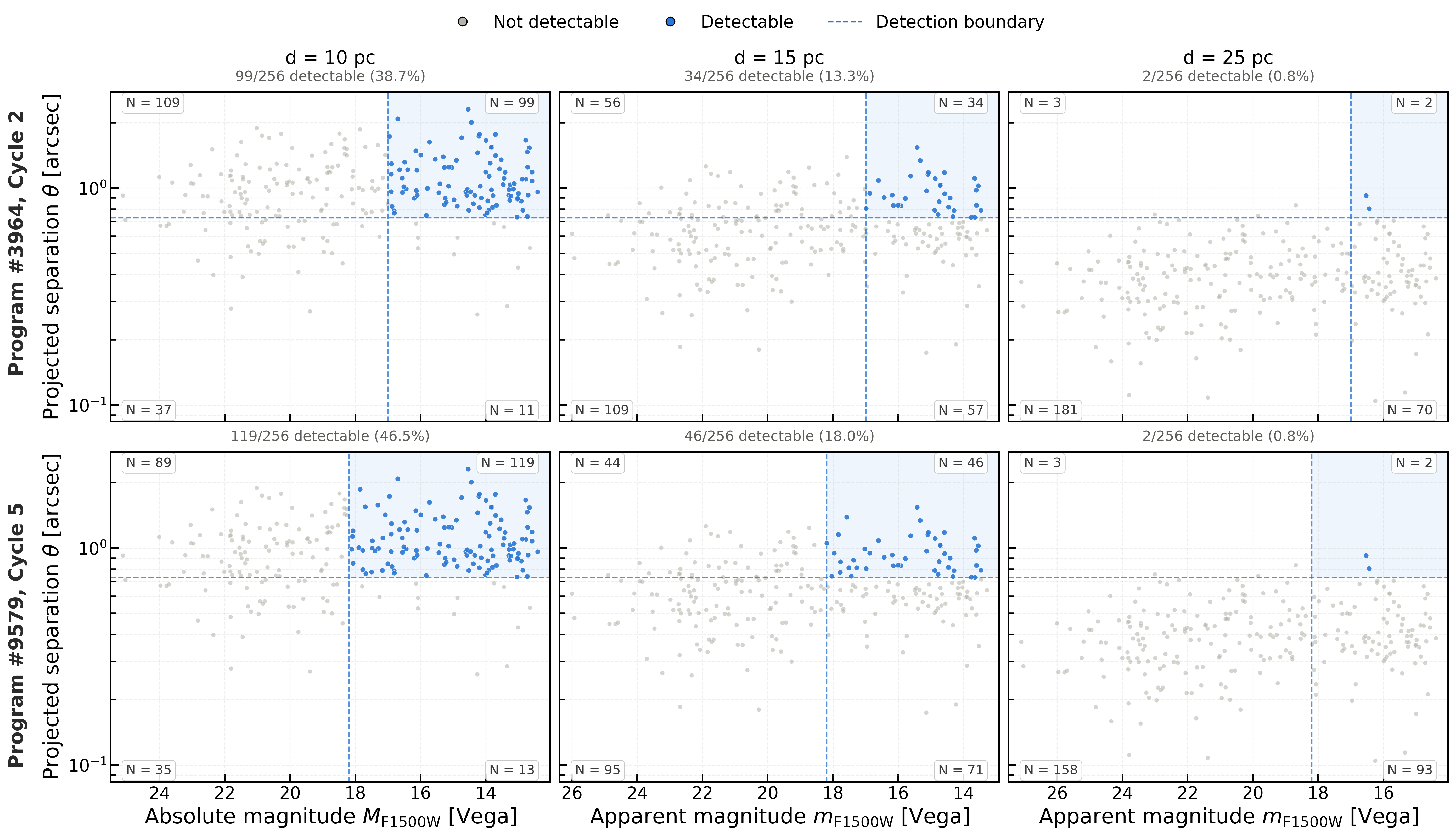}
    \end{minipage}
    \caption{The fraction of detectable (>5$\sigma$) substellar companions from our population synthesis through direct imaging for two different surveys performed with JWST as a function of distance. The upper panel corresponds to the MEAD survey \citep{albertetal26} while in the lower panel we assumed the observing strategy of The Fate of Jovian Worlds survey \citep{Mullally2026_JWST}. The detection limits for the brightness of the substellar companion and the projected angular resolution are given by the vertical and horizontal dotted lines, respectively. The fraction of detectable substellar companions decreases significantly with distance. While $38-46$~\% of the substellar companions at $10$~pc can be directly imaged, this number decreases to $13-18$~\% and only $0.8$\% for $15$ and $25$~pc, respectively. Clearly, the main limitation is the angular resolution since most companions are concentrated at relatively low separations.}
    \label{fig:di_dist}
\end{figure*}

\subsection{Detections predicted within the population synthesis}

To evaluate the detectability of substellar companions around WDs that descended from $1$--$4$~\Msun progenitors, we determined the present-day effective temperatures of the WDs and substellar companions generated by our population synthesis \citepalias{mauchsorianoetal26-1}. We restricted this analysis to the population of surviving substellar companions obtained with the weaker stellar tides proposed by \citet{Villaver_2009} and \citet{Mustill_2012}. 

Figure \ref{fig:twd_ssc} shows the resulting effective temperatures of the WDs and their substellar companions, together with the companion mass and temperature distributions. Most surviving substellar companions are gas giants with masses of $1$--$5$~\Mjup, and they are generally cold, with effective temperatures predominantly below $150$~K, reflecting the relatively old ages of the systems and the low masses of the companions. The WDs are also generally cool ($T\lesssim$~6000 K) and preferentially have relatively low masses ($M=0.53$--$0.66$~\Msun). The corresponding distribution of orbital separations is shown in Figure~4 of \citetalias{mauchsorianoetal26-1}.

To evaluate the detectability of these predicted substellar companions through direct imaging, we assigned a random orbital inclination and phase to each synthetic substellar companion and calculated the projected separation from the WD. We then combined the predicted radii and effective temperatures of both components with synthetic photometry in the JWST/MIRI F1500W filter. Adopting an exposure time of 444~s, as used by the MEAD survey, we determined whether each substellar companion would be sufficiently bright and separated from its host WD to be spatially resolved and detected. To assess how the sensitivity could impact the detectability of our population, we also employ the configuration from the Fate of Jovian Worlds survey \citep{Mullally2026_JWST} with an integration time of $t=3999$~s, which is designed to reach planet masses below half a Jupiter mass for all targets and below Saturn's mass for approximately 75\% of them. Integration times were converted into limiting magnitudes (and fluxes) using the JWST Exposure Time Calculator assuming medium background noise. 

\begin{figure}
    \centering
    \begin{minipage}{0.5\textwidth}
        \centering
        \includegraphics[width=1\textwidth]{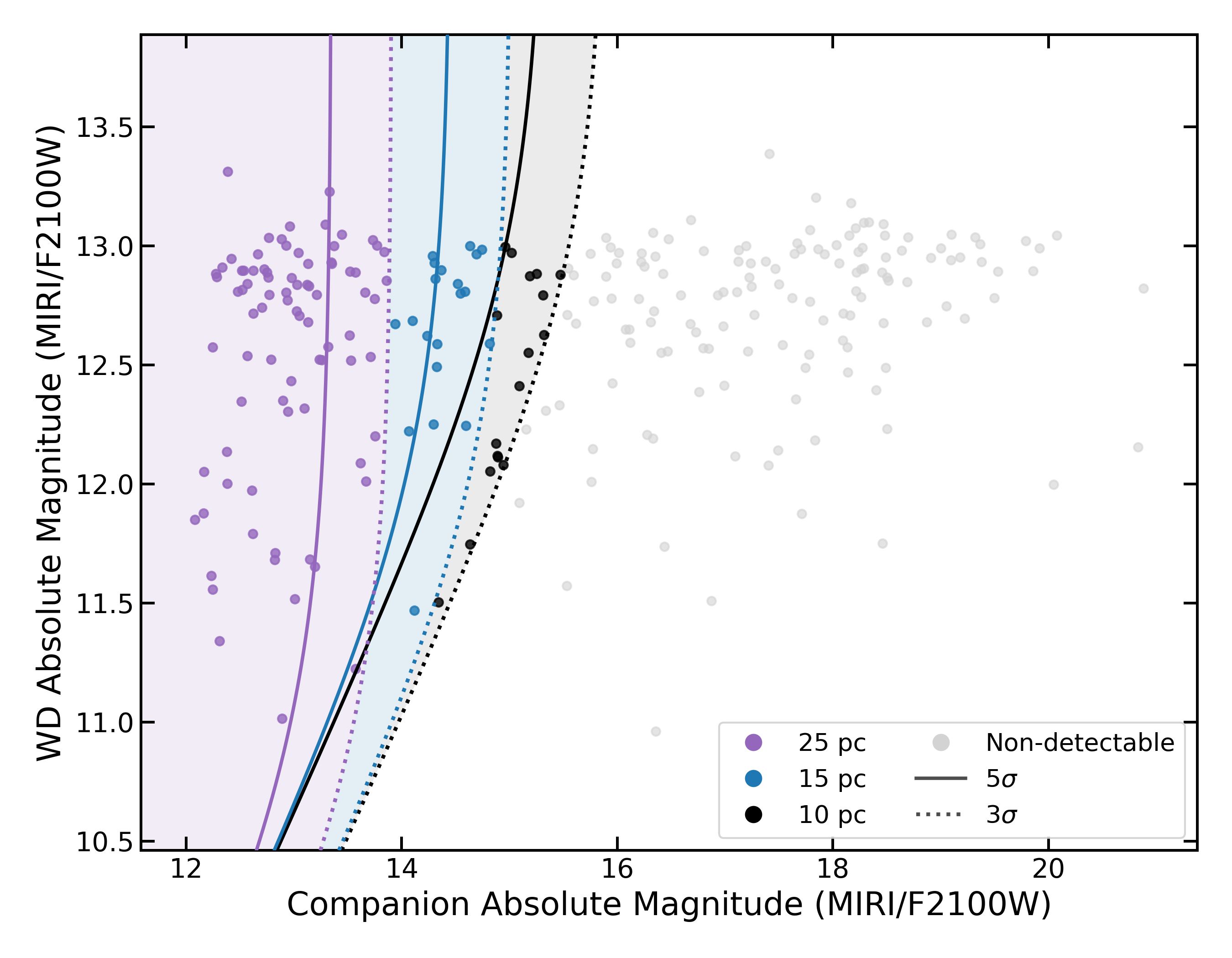}
    \end{minipage}
    \caption{Absolute magnitudes of the synthetic WD–substellar companion population. The curves show IR excess detectability thresholds at different distances assuming an exposure time of $t=710$~s. The IR excess detectability is evaluated under the assumption that all substellar companions remain spatially unresolved from their host WDs. At small distances (below 10 pc), a significant fraction ($\sim 40\%$) of the predicted substellar companion population can be detected through the IR excess method. However, at greater distances the detection is limited by sensitivity.}
    \label{fig:PS_IR}
\end{figure}

\begin{figure}
    \centering
    \begin{minipage}{0.5\textwidth}
        \centering
        \includegraphics[width=1\textwidth]{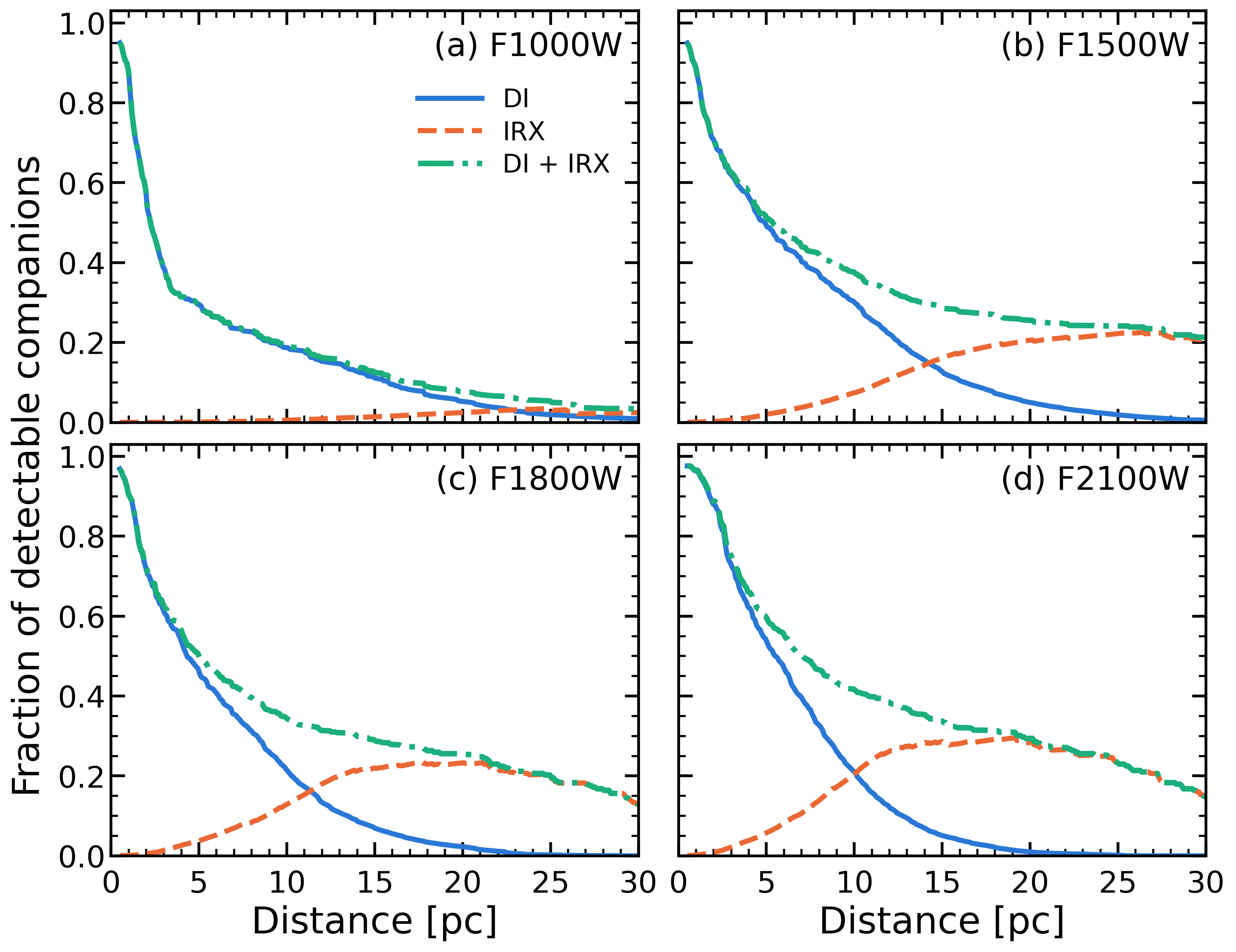}
    \end{minipage}
    \caption{Fraction of detectable substellar companions ($>5\sigma$) as a function of distance, considering both direct imaging (DI) and IR excess (IRX), for the four filters we consider (i.e. F1000W, F1500W, F1800W, and F2100W). The contribution of DI is significant at small distances but decreases rapidly with increasing distance, primarily because of the small angular separations between the substellar companions and their host WDs. For F1500W, F1800W, and F2100W filters, the fraction of substellar companions detectable through IRX exceeds that detectable through DI beyond approximately 10–15 pc. This transition occurs at distances for which the angular resolution of the respective filters correspond to projected physical separations comparable to the median orbital separation of our predicted substellar companion population (see Fig. 4 of \citetalias{mauchsorianoetal26-1}).}  
    \label{fig:detection_rate_vs_distance}
\end{figure}

Figure~\ref{fig:di_dist} shows the resulting distributions for systems located at distances of 10, 15, and 25~pc. For the adopted exposure time, the sensitivity of the observations corresponds to a limiting apparent magnitude of $m_{\text{Veg}}=17.0$ mag (for a $5\sigma$ detection), while the spatial resolution imposes a minimum projected angular separation of $\theta=0.732''$. As the distance increases, the substellar companions become fainter and their projected angular separations decrease, substantially reducing the fraction that can be resolved. If the entire predicted population were located at 10~pc, the adopted JWST configuration would resolve 38 to 46 \% of the surviving companions. This fraction decreases rapidly with distance. In particular, only 13 to 18 \% of companions would be detectable through DI at 15pc, where the main limitation is the angular resolution. At 30~pc none of the predicted companions would be detectable through spatially resolved imaging. 

The situation improves for detections through IR excess. Figure\,\ref{fig:PS_IR} illustrates that for the observing design 
chosen by the MEAD survey, and assuming all companions to be spatially unresolved,  45.4 \% of the predicted companions would be detected at a distance of $10$\,pc with $>3\sigma$ significance. This fraction decreases to $38.9$\% and $31.3$\% for $15$ and $25$\,pc, respectively. 
These fractions are, of course, not realistic because, as shown previously and depending on distance, a significant fraction of companions will be resolved. 
To estimate the total detection rate, IR excess detection should only be applied to unresolved companions.

To estimate the total fraction of the substellar companion population that would be detectable as a function of distance, we therefore combine both detection methods: for resolved companions, we analyze detectability under direct imaging, and for unresolved ones, we analyze detectability using IR excess. The results are shown in Figure \ref{fig:detection_rate_vs_distance}. At a distance of $30$\,pc, up to 
20\% of all companions will be detectable either through direct imaging or IR excess with current JWST surveys. Given that we predict only $\sim3\pm1.5$~\% of all WDs descending from $1$--$4$~\Msun~stars to host a substellar companion in the mass range of $0.8$--$30$~\Mjup, we expect only around $1\%$ of WDs at 30\,pc to host a detectable substellar companion.
 
However, using a fixed distance for our simulated population provides only a first guess. In the next section we use our results to estimate the detectability of companions around 
WDs within 30\,pc of the Sun.

\begin{figure*}
    \centering
    \begin{minipage}{1.0\textwidth}
        \centering
        \includegraphics[width=0.99\textwidth]{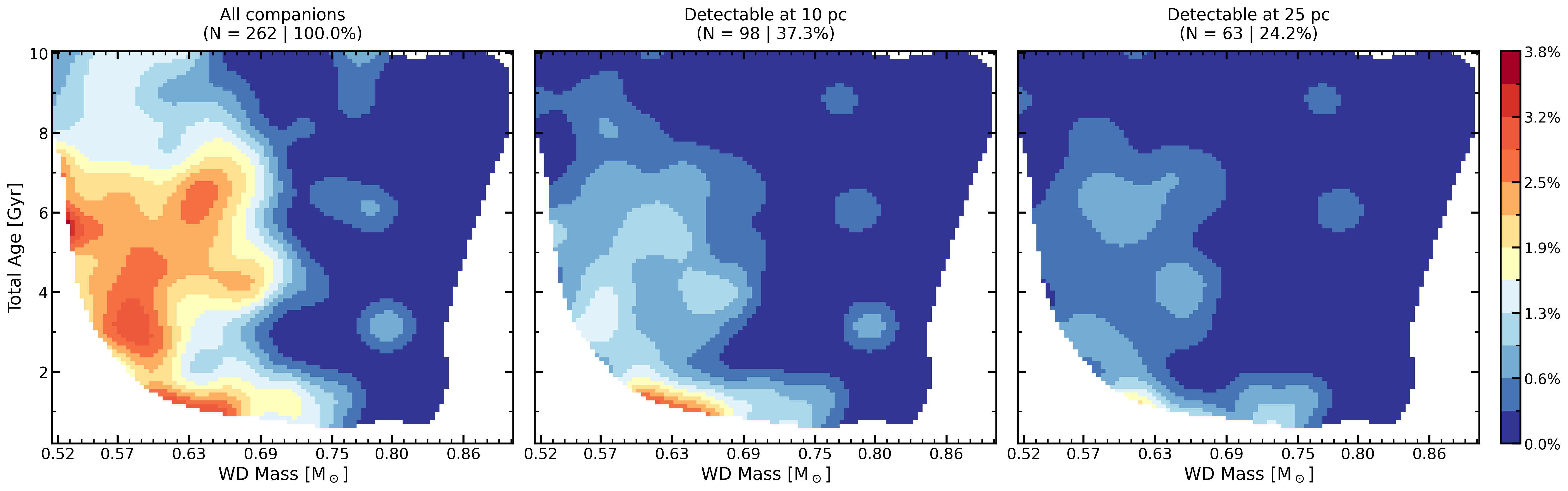}  
    \end{minipage}
    \caption{Left: Occurrence map showing the fraction of WDs hosting substellar companions as a function of total system age and WD mass. Center and Right: Detectability maps showing the fraction of WDs with detectable companions through direct imaging or IR excess at 10 and 25 pc, adopting a $5\sigma$ detection threshold in F1500W. These maps were computed using the bayesian formalism and provide continuous estimates of the detection probabilities based on the age and mass of a given WD. }
    \label{fig:PS_detectability_map}
\end{figure*}

\subsection{Predicted JWST yields in the solar neighbourhood}

 \begin{table}
    \caption{Selection criteria applied to the \citet{Gentilefusilloetal2021} WD catalogue.}
    \label{tab:wd_selection_criteria}
    \centering
    \begin{tabular}{lc}
        \hline\hline
        Parameter & Selection criterion \\
        \hline
        WD probability ($P_{\mathrm{WD}}$) & $P_{\mathrm{WD}} \geq 0.95$ \\
        WD mass ($M_{\mathrm{H}}$) & $0.50 \leq M_{\mathrm{WD}}/M_\odot \leq 0.85$ \\
        Effective temperature ($T_{\mathrm{eff,H}}$) & $1500 \leq T_{\mathrm{eff}}/\mathrm{K} \leq 90000$ \\
        Surface gravity ($\log g_{\mathrm{H}}$) & $7.9 \leq \log g \leq 9.3$ \\
        Distance ($r_{\mathrm{geo}}$) & $d < 30\,\mathrm{pc}$ \\
        \hline
    \end{tabular}
\end{table}
In order to assess the feasibility of detecting the predicted surviving substellar companions around real WDs, we employed the catalogue compiled by \cite{Gentilefusilloetal2021}, which contains $\sim1.3$~million WD candidates selected from Gaia Early Data Release 3 (EDR3). We then filtered our sample by imposing the conditions shown in Table \ref{tab:wd_selection_criteria}, that is, we selected high-confidence WDs within our mass range and with reasonable temperature and gravity values to ensure a clean, physically consistent sample suitable for the analysis performed with the \texttt{wdwarfdate} package described below. The selection resulted in $288$ WDs.

We used the \texttt{wdwarfdate} code \citep{kimanetal2022}, a publicly available Python package which employs Bayesian statistics to derive the cooling and total ages (from the zero-age main-sequence stage) of each WD based on its effective temperature ($T_{\mathrm{eff}}$) and surface gravity ($\log\,g$). For simplicity, we assumed that all WDs in our sample have DA spectral types (hydrogen-dominated atmospheres). The code uses cooling models from \cite{bedardetal2020} to estimate the WD masses and cooling ages. An initial-to-final mass relation (in our case, we adopted the model from \citealp[]{Cummingsetal2018}) was used to determine the mass of the progenitor star. With this value, the package uses MIST stellar evolution tracks \citep{dotteretal2016,choietal2016} to estimate the main-sequence lifetime and thus the total age of the system. 

As expected from a Bayesian model, each parameter obtained is represented as a probability density function. For our analysis, we adopted the median of each total age posterior distribution, along with the corresponding lower and higher uncertainty bounds (the 16th and 84th percentiles, respectively).
Additionally, we used the $T_{\mathrm{eff}}$ and $\log\,g$ values reported in the catalog to interpolate the synthetic WD spectra from \cite{Koester2010} and estimate the absolute magnitudes of each WD in the relevant JWST filters. These magnitudes were then used to evaluate the detectability of substellar companions through IR excesses.

To identify which companions in our synthetic population would be detectable around a given WD in the observed catalogue, we placed all synthetic systems at the distance of that WD. For spatially resolved companions, we evaluated their detectability based on their apparent magnitudes. For unresolved companions, we instead combined their predicted fluxes with those of the corresponding modelled WDs and assessed whether the resulting magnitudes in the relevant filters would produce a detectable IR excess. 

To determine the probability of a given WD having a detectable substellar companion based on our population synthesis, we considered, in addition to its distance, both its mass and the total age of the system. This is because the main factors determining the probability of a progenitor star hosting a substellar companion are stellar mass and metallicity \citep{wolthoff2022}. The latter, however, cannot be directly measured for observed WDs, as the progenitor stars' envelopes were expelled during the late stages of stellar evolution. However, in our sample, we used an age-metallicity relation that links stellar age to progenitor metallicity in our synthetic population.
We therefore compute the probability that a WD hosts a detectable substellar companion by constructing probability maps using Bayes' theorem, considering two continuous variables (WD mass and total age) and one discrete variable (detectable substellar companion, abbreviated det SC), that is \citep{Gregory_2005}:
\begin{equation}
\text{P}(\text{det SC} \mid M_{\mathrm{WD}}, t_{\text{tot}}) = \frac{f(M_{\mathrm{WD}}, t_{\text{tot}} \mid \text{det SC}) \cdot \text{P}(\text{det SC})}{f(M_{\mathrm{WD}}, t_{\text{tot}})}
\label{eq:Bayes_occ_2D}
\end{equation}
where $P(\mathrm{det\,SC}\mid M_{\mathrm{WD}},t_{\mathrm{tot}})$ is the posterior probability that a system with WD mass $M_{\mathrm{WD}}$ and total age $t_{\mathrm{tot}}$ hosts a detectable substellar companion (det SC); $f(M_{\mathrm{WD}},t_{\mathrm{tot}}\mid\mathrm{det\,SC})$ is the joint probability density function (PDF) of WD mass and total age for systems hosting a detectable substellar companion; $P(\mathrm{det\,SC})$ is the prior probability of a system hosting a detectable substellar companion in the synthetic population; and $f(M_{\mathrm{WD}},t_{\mathrm{tot}})$ is the joint marginal PDF of WD mass and total age across the full synthetic population.

This formulation provides a continuous estimate of the probability that a WD hosts a detectable substellar companion as a function of WD mass and total age, avoiding the need for arbitrary and rigid binning. Figure \ref{fig:PS_detectability_map} illustrates how the probability maps vary depending on distance.
The PDFs were estimated non-parametrically using kernel density estimation \citep{silverman_1986}. The kernel density estimators were implemented using the \texttt{KernelDensity} estimator of the  \texttt{scikit-learn} Python package \citep{Pedregosa_2011}, adopting its default Gaussian kernel function. Additionally, we used a common bandwidth for both KDEs, which was calculated using a generalization of Silverman's rule \citep{scott_1992}.

The mean probability of hosting a detectable substellar companion and its associated uncertainty, $\text{P}(\text{det SC}\mid M_{\mathrm{WD}},t_{\text{tot}}) \pm \sigma_P$, were estimated through Monte Carlo sampling of the uncertainties in the inferred WD mass and total age. For each WD, we generated 1000 realizations of the $(M_{\mathrm{WD}},t_{\text{tot}})$ pair by independently sampling from Gaussian distributions centered on their respective inferred values, with standard deviations corresponding to their estimated uncertainties.

Once we obtained the probability of hosting a detectable substellar companion for each WD, the total number of detectable substellar companions in the sample was estimated assuming independence between the stellar remnants. Under this assumption, the expected number of detectable substellar companions is given by the sum of the expected Bernoulli outcomes:
\begin{equation}
N_p = \mathbb{E}[X] = \sum_{i=1}^{N} p_i,
\label{eq:n_p}
\end{equation}
where \(p_i\) is the mean probability of hosting a detectable substellar companion for WD \(i\).

The uncertainty associated with this prediction was obtained from the intrinsic statistical dispersion arising from the Bernoulli nature of each detection trial. Accordingly, the total standard deviation is given by
\begin{equation}
\sigma_{N_p,\mathrm{total}} =
\sqrt{
\sum_{i=1}^{N} p_i(1-p_i)
}.
\label{eq:n_p_error}
\end{equation}

Table \ref{tab:detection_summary} shows the detectability results for the 288 WDs within 30pc. Only in F1000W does direct imaging yield a greater detection fraction than the IR excess ($5.47$\% compared to $3.18$\% for $5\sigma$). In contrast, the advantage of the IR excess method increases significantly towards longer wavelengths. In particular, around $20-30\%$ of the predicted planet population would be detectable through this method, whereas direct imaging would yield a very low detection fraction of just $4-7$\% of the population. Despite the promising rates of the former method, they translate into an absolute number of just 0.6-0.9 detected companions. The predicted low occurrence rate therefore explains this low value, given that in our baseline models we find that just $1.3\%$ of the WDs that descended from $1-4$~\Msun stars hostƒ a substellar companion.

We therefore predict that, given the population inferred from radial velocity surveys and the assumed age–metallicity relation, current JWST surveys will yield virtually no detections. In contrast, future surveys targeting more massive WDs may be more promising, because their substellar companion populations remain poorly constrained. WDs that descended from these stars provide an alternative opportunity to constrain this population. 

For instance, recent observations with ground-based facilities such as the Very Large Telescope (VLT), using the Enhanced Resolution Imager and Spectrograph (ERIS), have enabled the derivation of a $95\%$ upper limit of approximately $16\%$ on the occurrence rate of companions more massive than $\sim12\,\mathrm{M_{Jup}}$ at semi-major axes of $\approx50$--$100$\,au (Lagos-Vilches et al., A\&A~submitted). Their results are, however, limited by the field of view of their instruments, which could explain their non-detection: 

While gravitational instability can produce substellar companions across a broad range of stellar masses, its relative importance is expected to increase towards higher stellar masses ($M\gtrsim3.5$\Msun), where core accretion becomes increasingly inefficient. Companions formed through gravitational instability are generally expected to be relatively massive and to orbit at large separations. Given the orbital expansion due to stellar mass loss, these companions are predicted to be at very large separations from their WDs. In this context, JWST/MIRI observations could provide significant advantages compared to VLT/ERIS. In its widest imaging mode, MIRI covers a sky area approximately 12 times larger than the 13 mas/pixel configuration of ERIS/NIX, and about 2.7 times larger than its 27 mas/pixel configuration \citep{Davies_2023,Dicken_2024}.

\begin{table*}
\caption{Summary of detection results with JWST/MIRI imaging for the 288 WDs within 30\,pc, based on our population synthesis and the two most important variations.}
\label{tab:detection_summary}
\centering
\small
\setlength{\tabcolsep}{6pt}
\begin{tabular}{l l l c c c c}
\hline\hline
\noalign{\smallskip}
 & & & \multicolumn{2}{c}{$3\sigma$} & \multicolumn{2}{c}{$5\sigma$} \\
\cline{4-5} \cline{6-7}
\noalign{\smallskip}
Case & Filter & Method & $f_{\rm det}$ [\%] & $N_{\rm det}$ & $f_{\rm det}$ [\%] & $N_{\rm det}$ \\
\noalign{\smallskip}
\hline
\noalign{\smallskip}
Baseline & F1000W & DI & 6.80 & $0.23 \pm 0.48$ & 5.47 & $0.18 \pm 0.43$ \\
 &  & IRX & 5.17 & $0.14 \pm 0.38$ & 3.18 & $0.09 \pm 0.30$ \\
 &  & DI\,+\,IRX & 11.97 & $0.37 \pm 0.61$ & 8.65 & $0.27 \pm 0.52$ \\
 & F1500W & DI & 7.31 & $0.28 \pm 0.53$ & 6.64 & $0.26 \pm 0.51$ \\
 &  & IRX & 23.74 & $0.73 \pm 0.85$ & 20.05 & $0.61 \pm 0.78$ \\
 &  & DI\,+\,IRX & 31.05 & $1.01 \pm 1.00$ & 26.70 & $0.86 \pm 0.93$ \\
 & F1800W & DI & 4.58 & $0.18 \pm 0.42$ & 3.96 & $0.15 \pm 0.39$ \\
 &  & IRX & 24.40 & $0.76 \pm 0.87$ & 19.13 & $0.58 \pm 0.76$ \\
 &  & DI\,+\,IRX & 28.98 & $0.94 \pm 0.97$ & 23.09 & $0.73 \pm 0.86$ \\
 & F2100W & DI & 4.04 & $0.16 \pm 0.40$ & 3.58 & $0.14 \pm 0.37$ \\
 &  & IRX & 29.92 & $0.94 \pm 0.97$ & 23.21 & $0.71 \pm 0.84$ \\
 &  & DI\,+\,IRX & 33.96 & $1.10 \pm 1.05$ & 26.79 & $0.85 \pm 0.92$ \\
\noalign{\smallskip}
\hline
\noalign{\smallskip}
No tidal forces & F1000W & DI & 8.47 & $0.82 \pm 0.90$ & 7.35 & $0.71 \pm 0.84$ \\
 &  & IRX & 11.92 & $0.99 \pm 0.99$ & 8.84 & $0.75 \pm 0.87$ \\
 &  & DI\,+\,IRX & 20.39 & $1.81 \pm 1.34$ & 16.19 & $1.47 \pm 1.21$ \\
 & F1500W & DI & 7.06 & $0.74 \pm 0.85$ & 6.51 & $0.67 \pm 0.82$ \\
 &  & IRX & 37.61 & $3.15 \pm 1.76$ & 32.11 & $2.69 \pm 1.63$ \\
 &  & DI\,+\,IRX & 44.67 & $3.88 \pm 1.96$ & 38.62 & $3.36 \pm 1.82$ \\
 & F1800W & DI & 4.33 & $0.45 \pm 0.67$ & 3.88 & $0.40 \pm 0.63$ \\
 &  & IRX & 37.80 & $3.16 \pm 1.77$ & 31.11 & $2.60 \pm 1.60$ \\
 &  & DI\,+\,IRX & 42.13 & $3.60 \pm 1.88$ & 34.98 & $3.00 \pm 1.72$ \\
 & F2100W & DI & 3.62 & $0.37 \pm 0.60$ & 3.31 & $0.33 \pm 0.57$ \\
 &  & IRX & 45.02 & $3.83 \pm 1.94$ & 36.22 & $3.07 \pm 1.74$ \\
 &  & DI\,+\,IRX & 48.64 & $4.20 \pm 2.03$ & 39.53 & $3.40 \pm 1.83$ \\
\noalign{\smallskip}
\hline
\noalign{\smallskip}
Outer companion population & F1000W & DI & 35.92 & $11.17 \pm 3.27$ & 31.88 & $9.96 \pm 3.10$ \\
 &  & IRX & 1.56 & $0.46 \pm 0.68$ & 1.16 & $0.35 \pm 0.59$ \\
 &  & DI\,+\,IRX & 37.49 & $11.63 \pm 3.34$ & 33.04 & $10.31 \pm 3.15$ \\
 & F1500W & DI & 51.82 & $16.12 \pm 3.89$ & 47.79 & $14.82 \pm 3.74$ \\
 &  & IRX & 7.20 & $2.09 \pm 1.44$ & 6.51 & $1.89 \pm 1.37$ \\
 &  & DI\,+\,IRX & 59.02 & $18.21 \pm 4.12$ & 54.30 & $16.70 \pm 3.96$ \\
 & F1800W & DI & 44.91 & $14.15 \pm 3.66$ & 40.06 & $12.61 \pm 3.46$ \\
 &  & IRX & 9.47 & $2.78 \pm 1.66$ & 8.09 & $2.37 \pm 1.53$ \\
 &  & DI\,+\,IRX & 54.38 & $16.93 \pm 3.98$ & 48.14 & $14.99 \pm 3.76$ \\
 & F2100W & DI & 47.29 & $14.93 \pm 3.75$ & 41.50 & $13.08 \pm 3.52$ \\
 &  & IRX & 12.52 & $3.71 \pm 1.91$ & 10.57 & $3.12 \pm 1.76$ \\
 &  & DI\,+\,IRX & 59.81 & $18.64 \pm 4.16$ & 52.07 & $16.20 \pm 3.90$ \\
\noalign{\smallskip}
\hline
\end{tabular}
\tablefoot{%
We tested the impact of the assumed strength of tidal forces and of a potential companion population at large separations. Each entry shows the results for each filter through direct imaging (DI), infrared excess (IRX) and the combination of both (DI+IRX), at $3\sigma$ and $5\sigma$ significance. Columns correspond to the fraction of substellar companions that would be detected, $f_{\rm det}$, averaged over the 288 WDs within 30 pc, and the expected number of WDs with a detectable companion, $N_{\rm det}$, with its statistical uncertainty.
}
\end{table*}

\section{Discussion}

We have estimated that even JWST is unlikely to detect Jupiter-mass planets or low-mass brown dwarfs around WDs that descend from $1$--$4$~\Msun~progenitors. For already completed surveys such as the MEAD and MEOW surveys, our results explain the absence of detections. For ongoing surveys such as the Fate of Jovian Worlds survey our results predict small chances of successfully detecting Jupiter-mass planets while our simulations do not predict the occurrence rates or detectability of Saturn-mass companions. Furthermore, our predictions depend on several assumptions and observational constraints which could be subject to biases or selection effects. In what follows, we discuss each of these uncertainties and their implications on our results.    

\subsection{Age-metallicity relation} 

In our population synthesis presented in \citetalias{mauchsorianoetal26-1}, we adopted the global age--metallicity relation derived by \citet{carrillo2023}. However, recent work by \citet{Fernandez-Alvar-2025} provides evidence that the global age--metallicity relation of \citet{carrillo2023} may underestimate the metallicity of the solar neighborhood at early times. Based on fits to the Gaia colour--magnitude diagram, they infer significantly higher local metallicities at early ages. In \citetalias{mauchsorianoetal26-1}, we showed that adopting this relation as the enrichment history of the solar neighborhood increases the predicted substellar companion occurrence rate by approximately a factor of two and shifts the region of highest occurrence toward older ages.

This alternative age--metallicity relation therefore affects both the predicted occurrence rates and the age distribution of surviving companions. However, the larger number of companions predicted at older ages does not translate directly into a correspondingly larger number of detectable systems, because older companions are cooler and hence more difficult to detect through either direct imaging or IR excess. We therefore conclude that uncertainties in the age--metallicity relation introduce systematic uncertainties of less than a factor of two in our predicted occurrence and detection rates.

\subsection{Stellar tides and the survivability of companions} 

In \citetalias{mauchsorianoetal26-1}, we adopted two tidal prescriptions to determine which planets from our initial population survive the post-main-sequence evolution of their host stars: the relatively strong tides predicted by \citet{rasioetal96-1} and the weaker prescription adopted by \citet{Villaver_2009}. Recently, however, \citet{esseldeursetal26-1} showed that a more detailed consideration of the stellar envelope structure, combined with updated prescriptions for the frequency dependence of the turbulent viscosity, can substantially reduce the efficiency of tidal dissipation compared with the prescriptions of \citet{Villaver_2009} and \citet{Mustill_2012}, at least for the evolution of the Sun.

To assess the potential impact of overestimating tidal dissipation on our predictions, we considered the limiting case in which stellar tides are neglected entirely and simulated the orbital evolution of all potential WD companions accordingly. This provides an approximate upper limit on the fraction of companions that can survive the giant phases of their host stars. The predicted fraction of WDs with surviving substellar companions increases from 1.3\% to 3.2\% (from 262 to 652 surviving companions), while the expected number of WDs with a detectable companion increases by a larger factor of $\sim$4 (see the second case of Table \ref{tab:detection_summary}), since the additional surviving companions are more easily detected through IR excess. Thus, even in the extreme limit of no tidal forces, the expected number of detections remains small, increasing from $\sim$1 to $\sim$3--4.

\subsection{Improving direct-imaging detectability through kernel phase interferometry}

Recently, \citet{Adelman2025} presented the possibility of Kernel Phase Interferometry (KPI) for JWST/MIRI. Applying this technique to 16 white dwarfs from the MEOW survey, they reached $5\sigma$ contrasts of 5--6~mag at separations of $\sim200$~mas in F770W ($7.7\,\mu$m) for most targets. For this filter, the theoretical diffraction limit is $\lambda / D \approx 244$~mas which means that KPI allows to access the super-resolution regime. 

To test how the implementation of this technique could affect our results, we simply rescale the angular resolution for the filters we analyzed previously. 
The re-scalling factor depends on the effective wavelength, that is, $10/7.7$ (for F1000W), $15/7.7$ (for F1500W), $18/7.7$ (for F1800W), and $21/7.7$ (for F2100W). 

We find that around $45$ to $50\% $ of our synthetic substellar companion population would be resolvable at 25~pc, which is substantially higher than when considering traditional angular resolution limits (see Fig. \ref{fig:di_dist}). 
Moreover, when we apply the improved resolution to the 288 WDs within 30~pc, the fraction of companions detectable through direct imaging increases 
to 11.42\% under F2100W and reaches 18.18\% for F1500W (for a $>5\sigma$), which represents a significant improvement (see the fourth column of Tab. \ref{tab:detection_summary}). 
However, given the low predicted occurrence rate, even this improved resolution does not change our prediction of an expected number of less than one detectable substellar companion within 30~pc. 

\subsection{Deep observations}

Our baseline detection results are based on the observing configurations of two large, shallow surveys (MEAD and MEOW), with integration times ranging from 277 to 710~s. However, deeper observations will target $30$ WDs with F1500W for integration times of around \(t=3999\)~s have been approved \citep[GO 9579,][]{Mullally2026_JWST}. To explore how such an increase in exposure time affects our predictions, we adopted a similar observing setup in our simulations. Specifically, we consider 90 groups per integration, one integration per exposure, and 16 dithers for F1000W and F1500W. For F1800W and F2100W, we use 15 groups per integration, 6 integrations per exposure, and 16 dithers, resulting in a total integration time of \(t=4218\)~s.

In this scenario, 
the fraction of companions detectable through either IR excess or direct imaging
increases by a factor of around 1.1–1.4. In particular, for the F1500W, F1800W, and F2100W filters, we find that 35–42\% of the total population would be detectable within 30~pc, yielding a slightly higher expected number of detected companions (up to 1.3). We therefore conclude that even deep observational campaigns targeting planets around WDs are expected to yield very few detections of Jupiter-mass companions. However, our predictions cover only companions exceeding $0.8$~\Mjup.
If a substantial population of lower-mass companions, with masses closer to that of Saturn, exists, as suggested by recent theoretical studies \citep[][Ronco et al. A\&A~submitted]{Johnston2024}, longer exposure times could lead to a higher number of detections.

\subsection{Planets at large separations} 

Our baseline population synthesis is based on radial velocity surveys for giant planets around giant stars \citep{wolthoff2022}. Radial-velocity searches for exoplanets are heavily biased towards short orbital periods and it is therefore possible that a substellar companion population at large orbital separations exists around the progenitor stars of WDs.  

Indeed, based on a direct-imaging campaign of $300$ stars more massive than $1.5$~\Msun, \citet{nielsenetal19-1} derived an occurrence rate of $9^{+5}_{-4}\%$ for $5\!-\!13$\,\Mjup planets at separations of $10\!-\!100$~au. 
While this estimate is based on confirmed planets detected around just $7$ of the observed host stars, the result still indicates a population of companions beyond the detectability through radial velocities. 

Similarly, \citet{Vigan_2021} derived that the frequency of systems containing at least one substellar companion is $23.0_{-9.7}^{+13.5}$ and $5.8_{-2.8}^{+4.7}$ \% for BA and FGK stars, respectively. These rates are based on the detection of $16$ confirmed/previously known substellar companions, with 13 detections relevant to their statistical analysis, around $150$~observed stars and have been calculated for substellar companion masses between $1$ and $75$~\Mjup~and semimajor axes between $5$ and $300$~au. The existence of such planets at large orbital separations is in agreement with the predictions of most recent planet formation models around intermediate mass stars (Ronco et al. 2026, A\&A~submitted). 

At least part of the wide-orbit companion population that seems to exist according to imaging surveys \citep{Vigan_2021}, may have formed through gravitational instability and subsequent disk fragmentation. Our baseline population synthesis, which is based on radial-velocity surveys, may therefore not only underestimate the total occurrence rate of substellar companions but also preferentially sample companions formed through core accretion.

To test the impact of such wide companions on searches for companions around WDs, we randomly assigned substellar companions to our synthetic WD population based on the median occurrence rates reported by \citet{Vigan_2021}, splitting the sample by spectral type. Because their occurrence rates are valid up to $3$~\Msun stars, we simply assign companions within that progenitor mass range. We also adopted the mass functions for substellar companions as well as initial eccentricity and separation distributions from \citet{Vigan_2021}. We excluded companions with masses exceeding $30$~\Mjup. As tides are unlikely to affect these wide companions (see Fig. 3 of \citetalias{mauchsorianoetal26-1}), we simply assumed adiabatic expansion due to stellar mass loss during the late stages of stellar evolution. This test resulted in a global occurrence rate of $1916/20315$ ($9.4$\%) WDs hosting wide-orbit companions, characterized by orbital separations with a median of $57^{+174}_{-34}$au. 

To test the detectability of this population, we followed the procedure described in the previous sections and found that around $40-50\%$ of this population would be detectable through the direct imaging in F1500W, F1800W and F2100W filters (see Tab.~\ref{tab:detection_summary} for the exact numbers). This would correspond to a total of up to $\sim16$ detectable companions within 30pc (and up to $\sim$19 when combining both methods).

\section{Conclusions}

Based on the results of radial velocity surveys of substellar companions around giant stars \citep{wolthoff2022}, we calculated the detectability of substellar companions around WDs with JWST, considering companions that survived the evolution of their host stars. In a complementary study, we showed that the number of companions surviving common-envelope evolution is negligible \citep{mirandaetal2026}. 

Our results predict that JWST will detect only $\sim1$ substellar companion around WDs that descended from $1-4$~\Msun~stars within 30\,pc, even if the entire WD population within this volume were observed. Accounting for uncertainties in the age–metallicity relation and the strength of tidal interactions during the giant phases of the host stars could increase the expected number of detections to a few companions. Thus, if radial-velocity surveys of $1-4$~\Msun~giant stars \citep{wolthoff2022} provide a reasonably complete census of their substellar companions, searches around their descendants ($\sim0.5-0.85$~\Msun WDs) would have very limited prospects for success, even with JWST.

One possible caveat is that radial-velocity surveys are insensitive to companions at large orbital separations ($P\gtrsim3600$ d). Indeed, \citet{Vigan_2021} inferred occurrence rates exceeding $10$\% for substellar companions around intermediate-mass stars from high-contrast imaging surveys. Adopting these substantially higher, albeit uncertain, occurrence rates could yield nearly 20 detections if all WDs within 30~pc were observed with JWST.

These results suggest that the detection of substellar companions formed through core accretion around WD descendants will remain limited. This formation channel is expected to dominate planet formation around stars with masses below $\sim4$~\Msun. The combination of low occurrence rates and small detection probabilities contributes to this limitation. However, JWST observations of WDs may offer an opportunity to address a potentially more fundamental question: the formation of substellar companions around more massive stars.

Although gravitational instability may contribute to the population of wide substellar companions across the entire progenitor mass range considered here, its relative importance may increase substantially for stars more massive than $\sim4$~\Msun, around which giant planet formation through core accretion becomes increasingly inefficient. 

Substellar companions formed through gravitational instability and subsequent fragmentation around $\sim4-8$~\Msun~stars are difficult to detect during the main-sequence phase, whether through radial velocities, transits, or direct imaging. Their WD descendants, however, offer considerably more favorable conditions for detection. The short main-sequence lifetimes of these progenitors imply relatively young and therefore brighter substellar companions, while the substantial stellar mass loss during post-main-sequence evolution leads to greater orbital expansion and larger angular separations. Interestingly, these companions might be located at separations too large to be detected with ground-based and small field of view instruments such as VLT/ERIS (Lagos-Vilches et al. 2026, A\&A~submitted). Consequently, JWST surveys targeting WDs that descended from $\sim4-8$~\Msun~stars could provide a census of these substellar companions and probe the possible transition between dominant substellar companion-formation mechanisms as a function of progenitor stellar mass.


\begin{acknowledgements}
AMS acknowledges support from Agencia Nacional de Investigación y Desarrollo (ANID) Becas/Doctorado Nacional 21263654.
MRS thanks for support from the research unit FOR2990 (eRO-STEP: Stellar endpoints with eROSITA) funded by the Deutsche Forschungsgemeinschaft and the internal research project PI\_LIR\_26\_07 at USM.
MPR is partially supported by PIP 1122025010 0965CO from CONICET (Argentina). FLV acknowledges support from ANID FONDECYT Postdoctoral Grant No. 3250464. DB thanks for support from the National Natural Science Foundation of China (NSFC, Nos. W2632018, 12288102, 12125303, and 12525304), the Strategic Priority Research Program of the Chinese Academy of Sciences (grant Nos. XDB1160303, XDB1160300, XDB1160000, XDB1160200, XDB1160201), the CAS Project for Young Scientists in Basic Research (YSBR-148), the Yunnan Revitalization Talent Support Program ``YunLing Scholar'' project, International Centre of Supernovae (ICESUN), Yunnan Key Laboratory of Supernova Research (No. 202505AV340004),  the New Cornerstone Science Foundation through the XPLORER PRIZE, and the Yunnan Revitalization Talent Support Program-Science \& Technology Champion Project (No. 202305AB350003). This research has made use of the SVO Filter Profile Service "Carlos Rodrigo", funded by MCIN/AEI/10.13039/501100011033/ through grant PID2023-146210NB-I00
\end{acknowledgements}

\bibliographystyle{aa} 
\bibliography{WDplanets.bib}

\begin{appendix}





\FloatBarrier 
\clearpage

\end{appendix}
\end{document}